\documentclass[twocolumn,showpacs,preprintnumbers,draftclsnofoot]{revtex4}
\usepackage{amsmath}
\usepackage{amssymb}
\usepackage{graphicx}
\usepackage{dcolumn}
\usepackage{bm}
\usepackage{amssymb}
\usepackage{graphicx}
\usepackage{dcolumn}
\usepackage{bm}
\begin{document}

\title{ Light storage in wavy dielectric grating with Kerr nonlinearity  }
\author{Ma Luo\footnote{Corresponding author:swym231@163.com} }%luoma@gpnu.edu.cn,swym231@163.com
\affiliation{School of Optoelectronic Engineering, Guangdong Polytechnic Normal University, Guangzhou 510665, China}

\begin{abstract}

Periodical corrugation in the dielectric slab transforms the two waveguide modes at zero Bloch wave number into a leaky resonant mode and symmetry-protected bound states in the continuum (BIC) with a small frequency detune. The leaky resonant mode can be directly excited by a weakly linearly polarized normally incident optical field. In the presence of Kerr nonlinearity, BIC can be indirectly excited by an optically bistable response. Two types of bistable operations were considered. For the first type, the intensity of the incident field was gradually increased to exceed the critical value  and then decreased to zero. For the second type, the intensity was fixed, while the linear polarization angle of the incident field was gradually increased to exceed a critical value and then decreased to 0$^{o}$. Theoretically, the indirectly excited BIC can store  optical energy without loss, even when the intensity of the incident field decreases to zero. The incidence of an optical field with double frequency or orthogonal linear polarization can erase the stored optical field by destroying BIC. The proposed system can function as an optical storage and switching device.

\end{abstract}

\pacs{00.00.00, 00.00.00, 00.00.00, 00.00.00}
\maketitle

\section{Introduction}

Kerr optical materials exhibit a third-order nonlinear optical effect, whose reflective index is a linear function of the intensity of the optical field \cite{KerrMat1,KerrMat2}. As the power of the incident optical field changes, the optical response of photonic structures consisting of Kerr materials also changes, owing to the change in the relative permittivity. The Kerr effect in the bulk is weak because the third-order susceptibility is small. Photonic structures, such as photonic crystals \cite{pc1,pc2,pc3,pc4,pc5,pc6,pc7,pc8,pc9}, dielectric gratings \cite{fengwu19,fengwu21}, metallic thin films \cite{FZhou10,DavidSmith15}, semiconductor thin films \cite{semi1,semi2,semi3,semi4,semi5}, and graphene \cite{Christensen15}, can enhance the local density of states of the optical field, which in turn enhances the Kerr effect. Optical bistability is one of the most extensively studied Kerr effects \cite{bist1,bist2,bist3,bist4,bist5,bist6,bist7,bist8}. For a given incident power,  bistable optical devices can exhibit two bistable states with different optical responses. Optical devices can  switch between two bistable states as the incident power increases or decreases beyond a critical value \cite{pc8,semi2,semi3,Christensen15,switch1,switch2,switch3,switch4,switch5,switch6}. These two bistable states can be considered as ON and OFF states for all-optical switching devices in an integrated photonic system.

The mechanism of optical bistability in photonic structures is based on the excitation of the optical resonant mode with a Lorentz line shape \cite{FZhou10,DavidSmith15}, whose peak wavelength is approximately a linear function of the refractive index of  Kerr materials. The refractive index of  Kerr materials is a linear function of the intensity of the local optical field (i.e., $|\mathbf{E}|^{2}$), such that the peak wavelength is a linear function of $|\mathbf{E}|^{2}$. For a given power of the incident field, $|\mathbf{E}|^{2}$ is proportional to the incident power and Lorentz function at the incident wavelength. Because the Lorentz function is inversely proportional to the quadratic function of the difference between the peak wavelength and the incident wavelength, the self-consistent funcion gives a cubic function of $|\mathbf{E}|^{2}$. If the value of the incident wavelength minus the peak wavelength is a decreasing function of $|\mathbf{E}|^{2}$, the local optical field pulls the peak wavelength toward the incident wavelength, so that the cubic function can have three non-degenerated solutions. Two of the three solutions were in a stable state.  Bistable solutions exist if the difference between the incident wavelength and the peak frequency with $|\mathbf{E}|^{2}\approx0$ is larger than a critical value and the incident power is between two critical values. In other words, when the incident power is smaller (larger) than the lower (higher) critical value, $|\mathbf{E}|^{2}$ has only one branch of the solution that is connected to the lower (higher) branch of the bistable solutions. As the incident power increases (decreases) across the higher (lower) critical value, $|\mathbf{E}|^{2}$ switches from a lower (higher) branch to a higher (lower) branch. Optical responses, such as reflectance and transmittance, were proportional to $|\mathbf{E}|^{2}$. The optical resonant mode can be a leaky resonant mode \cite{FZhou10} or a quasi-BIC \cite{qbic1}.

In this study, we consider a situation in which a leaky resonant mode is nearly degenerated with a symmetry-protected bound state in the continuum (BIC), which occurs in the wavy dielectric grating \cite{maluo22} with small corrugation under a normal incident plane wave. The incident frequency was smaller than the resonant frequency of the BIC, and the resonant frequency of the BIC was smaller than that of the leaky resonant mode. At a low incident power, only the leaky resonant mode is excited, and the BIC cannot be excited by the incident plane wave owing to the symmetry mismatch. As the incident power increases across the higher critical value, the solution of $|\mathbf{E}|^{2}$ switches to the higher branch. In this case,  BIC can be excited by the scattered field owing to the modification of the spatial distribution of the refractive index. As the incident power subsequently decreases across the lower critical value, the energy in the leaky resonant mode is released, whereas the energy in the BIC remains  stored. As the incident power decreases further, the energy is still stored in the BIC without loss because the BIC is not coupled with the radiative mode. Thus, the local optical field does not decrease, and the refractive index remain modified. As the incident power reaches zero, the self-consistent solution of the BIC of the system with the modified refractive index is still lossless. To release the stored energy in the BIC, an incident optical field with different frequency or polarization is applied to destroy the symmetry mismatching between the modified refractive index and  BIC. This bistable procedure can be applied to light storage devices.

Another type of bistable procedure with a fixed incident power was considered, which was driven by tuning the linear polarization angle. In the previous case, both the leaky resonant mode and  BIC were TE-polarized. At the same incident frequency, the TM-polarized wave is far  from any resonant mode. Thus, the incident TM-polarized plane wave induces a weak local optical field. If the linearly polarized angle of the incident plane wave is between the TE and TM polarizations, only the TE component can excite a large local optical field and induce a significant Kerr effect. Thus, the effective incident power is given by the TE component of the incident field. Consequently, tuning the linearly polarized angle is equivalent to tuning the effective incident power. The effective incident power is smaller than the incident power; therefore, to switch to the higher branch of the bistable solution, the incident power must be larger than the higher critical value. After exceeding the higher critical value, BIC is excited. When the linearly polarized angle is further tuned, if the effective incident power remains nonzero, the modification of the refractive index is maintained so that the energy remains  stored in the BIC. When the linearly polarized angle is exactly equal to the TM polarization, the effective incident power is zero. In this case, the TM-polarized incident field induces the spatial distribution of the refractive index, which destroys the symmetry mismatching between the modified refractive index and BIC, which in turn releases the stored energy in the BIC.

The remainder of this paper is organized as follows: In Sec. II, the linear response of the wavy dielectric grating was discussed. In Sec. III, the bistable procedure of tuning the incident power is discussed. In Sec. IV, the bistable procedure of tuning the linearly polarized angle is discussed. In Sec. V, the conclusions are presented.

\section{Leaky resonant mode and symmetry-protected BIC of wavy dielectric grating }

\begin{figure}[tbp]
\scalebox{0.6}{\includegraphics{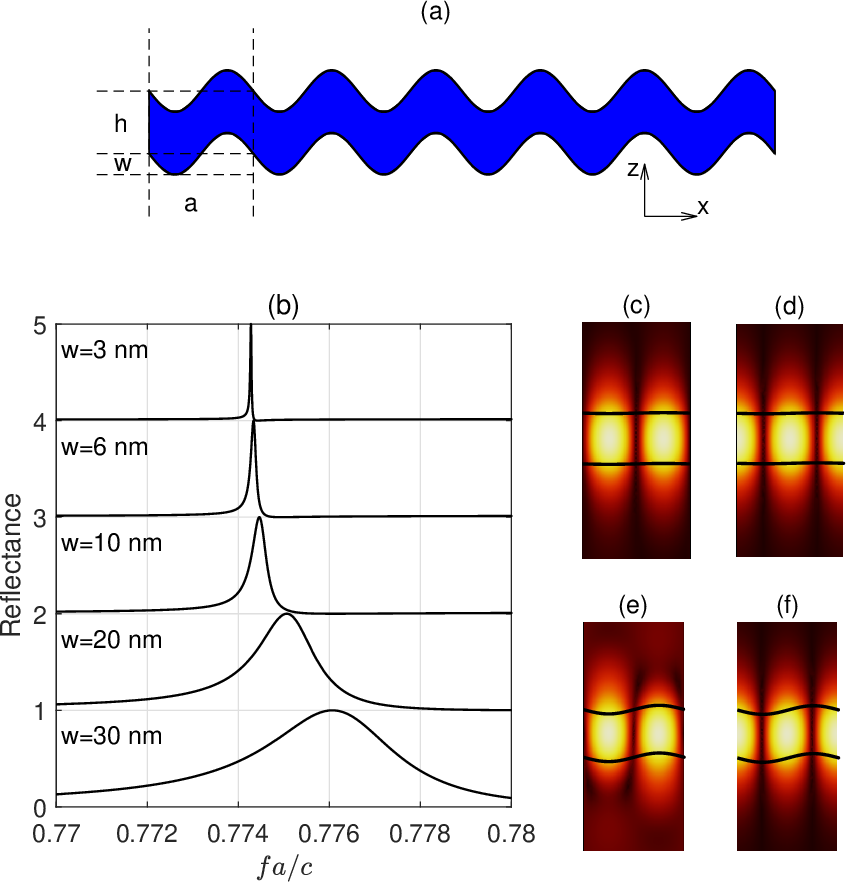}}% Here is how to import EPS art
\caption{ (a) Structure of the wavy dielectric grating. (b) TE reflectance spectra around $fa/c=0.775$ of the wavy dielectric grating for different values of $w$ under normal incidence. The field pattern of $|\mathbf{E}|$ of the leaky resonant modes and the symmetry-protected BICs are plotted in (c,e) and (d,f), respectively. In (c,d) and (e,f), $w=3$ nm and $w=30$ nm, respectively. The resonant frequency of the modes in (c-f) are $0.774279c/a$, $0.774263c/a$, $0.776241c/a$, and $0.774668c/a$, respectively. The Q factor of the modes in (c-f) are $2.3\times10^{4}$, $\infty$,  $2.3\times10^{2}$, and $\infty$, respectively. }
\label{figure_trans}
\end{figure}

The structure of the wavy dielectric grating is illustrated in Fig. \ref{figure_trans}(a). A dielectric slab with finite thickness along z-direction is periodically corrugated along x-direction. The thickness of the original dielectric slab was $h$. The structure is uniform along y-direction and has discrete translational symmetry along x-direction, with the period being $a$. The top and bottom surfaces of the wavy dielectric grating are defined by the function $z=\pm h/2-w\sin(2\pi x/a)$, where $w$ is the corrugation amplitude. A linearly polarized plane wave is normally incident on the wavy dielectric grating from z-direction. We consider TE polarization with the electric field along the y-direction. In the region with $z\gg0$, the electric field of the incident wave in the frequency domain is given as $\mathbf{E}_{inc}=\hat{y}E_{inc}e^{-i2\pi fz/c}$, where $f$ and $E_{inc}$ are the frequency and amplitude of the incident plane wave, respectively, and $c$ is the speed of light. If $f<c/a$, the reflected and transmitted fields in the regions with $z\gg0$ and $z\ll0$ are $\mathbf{E}_{ref}=\hat{y}E_{inc}e^{i2\pi fz/c}$ and $\mathbf{E}_{tra}=\hat{y}E_{inc}e^{-i2\pi fz/c}$, respectively. For a specific example of the numerical simulation, the parameters were assumed to be $a=800$ nm, $h=380$ nm, and $w$  between $3$ nm and $30$ nm. Air was used as the background medium. The relative permittivity of the dielectric medium with Kerr nonlinearity is assumed to be $\varepsilon_{r}=2.2+\chi^{(3)}|\mathbf{E}|^{2}$, where $\chi^{(3)}=4.4\times10^{-18}$ m$^{2}$/V$^{2}$, and $|\mathbf{E}|$ is the magnitude of the electric field \cite{KerrMat1,KerrMat2}. Materials with sufficient large value of $\chi^{(3)}$ can be obtained by doping the semiconductors or engineering the conjugation length of polymers \cite{kerrmat3}. The incident power, designated as $I_{inc}$, is proportional to the intensity of the incident electric field, which is given by $I_{inc}=\frac{1}{2}\sqrt{\frac{\varepsilon_{0}}{\mu_{0}}}|\mathbf{E}_{inc}|^{2}$. The electromagnetic field was numerically calculated by solving the two-dimensional Helmholtz equation of the y-component electric field, which is given as
\begin{equation}
\nabla\times\{\nabla\times[E_{y}(x,z)\hat{y}]\}-(\frac{2\pi f}{c})^{2}\varepsilon_{r}(x,z)E_{y}(x,z)\hat{y}=0\label{helmholtzE}
\end{equation}
where $\nabla=\hat{x}\frac{\partial}{\partial x}+\hat{z}\frac{\partial}{\partial z}$, $E_{y}=\mathbf{E}\cdot\hat{y}$. The spectral element method (SEM) \cite{SEM1,SEM2,SEM3,SEM4,SEM5,SEM6} was used to obtain a solution with high accuracy and efficiency. The computational domain was selected as a rectangular region of one period in the $\hat{x}-\hat{z}$ plane with $y=0$. Periodic boundary conditions are applied to the left and right boundaries. The top and bottom boundaries were truncated by a perfectly matched layer (PML). The total field-scattered field formula was applied to simulate incident plane waves. The self-consistent solution was obtained by iteratively calculating $E_{y}(x,z)$ and $\varepsilon_{r}(x,z)$.

When $I_{inc}$ is small, the nonlinear term in $\varepsilon_{r}$ can be neglected. Thus, we first discuss the linear response of the wavy dielectric grating without the Kerr nonlinearity. A flat dielectric slab with $w=0$ host waveguide modes whose dispersion relations are under the light cone. When a periodic boundary condition is imposed on the system, the dispersion relations are folded into the first Brillioun zone,  forming a band structure. Band crossings occur at the $\Gamma$ point of the first Brillioun zone, where the Bloch wave number is zero. The modes at the $\Gamma$ point were two-fold degenerate. Although the modes at the $\Gamma$ point are above the light cone, they are nonradiative because they are in the waveguide mode. When a periodic perturbation appears, that is, $w\ne0$, the band structure is modified and the degeneracy of each pair of modes at the $\Gamma$ point is broken. One of the modes is coupled with the radiative traveling waves ($\mathbf{E}_{ref}$ and $\mathbf{E}_{tra}$), so that the mode becomes a leaky resonant mode. The other mode is symmetry-mismatched with the radiative traveling wave, such that the mode is not coupled with any radiative traveling wave. The latter mode is the symmetry-protected BIC, which cannot be directly excited by a normally incident plane wave \cite{LeeJoannopoulos12}. The resonant frequencies of the two modes were slightly different. Under normal incident TE waves with a low intensity, the reflectance spectrum has a resonant peak around the resonant frequency of the leaky resonant mode. As $w$ decreased, the full width at half maximum (FWHM) of the peak decreased, as shown in Fig. \ref{figure_trans}(b). Thus, the Q factor of the leaky resonant mode is inversely proportional to $w$. At the resonant frequency, the magnitude of the electric field within the dielectric medium is enhanced owing to  excitation of the leaky resonant mode. The enhancement factor is inversely proportional to $w$.

In the absence of an incident field, the resonant modes can be calculated by solving the eigenvalue problem in Eq. \ref{helmholtzE}. The leaky resonant modes and  BICs of the system with $w=3$ nm and $w=30$ nm were calculated. The corresponding field patterns are plotted in Fig. \ref{figure_trans}(c-f) with the resonant frequencies and the Q factors provided in the caption. As $w$ increases from $3$ nm to $30$ nm, the Q factor of the leaky resonant mode decrease from $2.3\times10^{4}$ to $2.3\times10^{2}$. The field patterns of the leaky resonant modes and  BICs have nodal lines along y-direction at $x=0$($x=a/2$) and $x=a/4$($x=3a/4$), respectively, where $|\mathbf{E}|$ is zero. For the leaky resonant modes, the points with the maximum value of $|\mathbf{E}|$ are located at $x=a/4$($x=3a/4$) where $|\sin(2\pi x/a)|=1$, as shown in Fig. \ref{figure_trans}(c,e), such that the adjacent points have staggered $y$ coordinates. This feature induces scattering by the boundary of the wavy dielectric slab such that the Q factors are finite. For the BICs, the points with the maximum value of $|\mathbf{E}|$ are located at $x=0$($x=a/2$) where $|\sin(2\pi x/a)|=0$, as shown in Fig. \ref{figure_trans}(d,f), such that the points have the same $y$ coordinates. The energy only flux along $\pm$x-direction, but  not  along the y-direction, so that the energy loss is absent.

\begin{figure}[tbp]
\scalebox{0.58}{\includegraphics{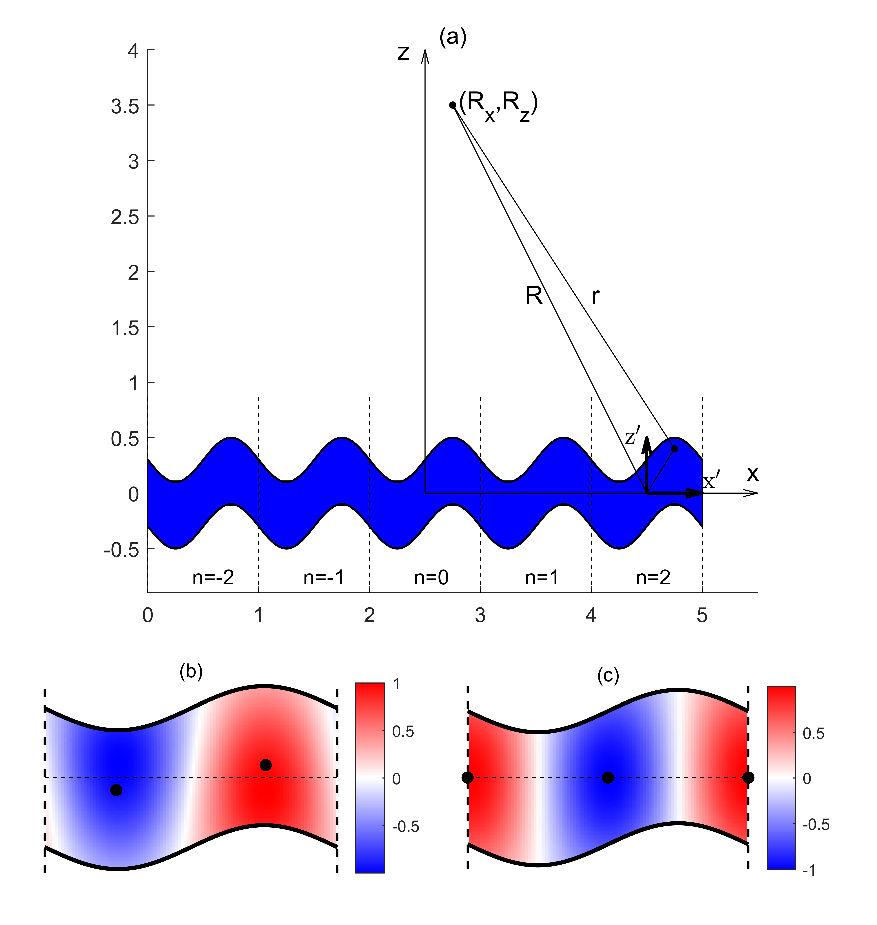}}% Here is how to import EPS art
\caption{ (a) The sketch map of the far-field calculation of the radiation field from the electric polarization vector of a resonant mode. (b) and (c) are the spatial distribution of $P_{y}$ of the leaky resonant mode and BIC, respectively. The black dots mark the location with maximum value of $|P_{y}|$. }
\label{figure_farfield}
\end{figure}

The symmetry-mismatch between BIC and the radiative mode can be interpreted by calculating the far-field radiation of the polarization current in the dielectric slab. The radiated field from a resonant mode can be calculated by integrating the radiation from polarization current within the dielectric slab. At an observation point $\mathbf{R}=R_{x}\hat{x}+R_{z}\hat{z}$, the vector potential of the radiated field from the n-th unit cell, $A_{y}^{n}$, is given as
\begin{equation}
A_{y}^{n}(\mathbf{R})=\frac{\mu_{0}}{4\pi}\int_{V_{n}}\frac{J_{P,y}(\mathbf{r}^{\prime})e^{ikr}}{r}d\mathbf{r}^{\prime}
\end{equation}
, with $J_{P,y}=-i2\pi fP_{y}$ being the polarization current and $P_{y}=(\varepsilon_{r}-1)\varepsilon_{0}E_{y}$ being the electric polarization vector within the dielectric slab at the local coordinate $\mathbf{r}^{\prime}=x^{\prime}\hat{x}^{\prime}+z^{\prime}\hat{z}^{\prime}$, as shown in Fig. \ref{figure_farfield}(a). The integral cover the dielectric slab region within the n-th unit cell designated as $V_{n}$, and $r=|\mathbf{R}-\mathbf{r}^{\prime}|=\sqrt{(na+x^{\prime}-R_{x})^2+(R_{z}-z^{\prime})^2}$. Considering the far-field limit with $R_{z}\gg a$, so that $r\approx R_{n}-e_{x}^{n}x^{\prime}-e_{z}^{n}z^{\prime}$ with $R_{n}=\sqrt{(na-R_{x})^2+(R_{z})^2}$, $e_{x}^{n}=(R_{x}-na)/R_{n}$ and $e_{z}^{n}=R_{z}/R_{n}$, the integral can be approximated as
\begin{equation}
A_{y}^{n}(\mathbf{R})=\frac{\omega\mu_{0}e^{ikR_{n}}}{4\pi R_{n}}(-i\langle P_{y}\rangle-ke_{x}^{n}\langle xP_{y}\rangle-ke_{z}^{n}\langle zP_{y}\rangle+\cdots)
\end{equation}
, with $\langle P_{y}\rangle=\int_{V_{n}}P_{y}(\mathbf{r}^{\prime})d\mathbf{r}^{\prime}$, $\langle xP_{y}\rangle=\int_{V_{n}}x^{\prime}P_{y}(\mathbf{r}^{\prime})d\mathbf{r}^{\prime}$, $\langle zP_{y}\rangle=\int_{V_{n}}z^{\prime}P_{y}(\mathbf{r}^{\prime})d\mathbf{r}^{\prime}$. Because of the periodic condition, $\langle P_{y}\rangle$, $\langle xP_{y}\rangle$ and $\langle zP_{y}\rangle$ are independent of $n$. The total vector potential is given as $A_{y}=\sum_{n=-\infty}^{+\infty}A_{y}^{n}$. The spatial distributions of $P_{y}$ for the leaky resonant mode and BIC are plotted in Fig. \ref{figure_farfield}(b) and (c), respectively. Because the structure of the wavy dielectric grating have inversion symmetry about the point at $(x^{\prime},z^{\prime})=(0,0)$, and mirror symmetric about the vertical axis with $x^{\prime}=\pm a/4$, the patterns of $P_{y}$ have the same symmetries. For the leaky resonant mode, $P_{y}$ is odd under the inversion transformation $(x^{\prime},z^{\prime})\rightarrow(-x^{\prime},-z^{\prime})$, so that $\langle P_{y}\rangle$ equates to zero. $xP_{y}$ and $zP_{y}$ are even under the inversion transformation, so that $\langle xP_{y}\rangle$ and $\langle zP_{y}\rangle$ are nonzero. Thus, the radiation field is nonzero. By contrast, for the BIC, $P_{y}$ is even under the inversion transformation $(x^{\prime},z^{\prime})\rightarrow(-x^{\prime},-z^{\prime})$, then $xP_{y}$ and $zP_{y}$ are odd under the inversion transformation, so that $\langle xP_{y}\rangle$ and $\langle zP_{y}\rangle$ equate to zero. In the left (right) half of the unit cell with $-a/2<x^{\prime}<0$ ($0<x^{\prime}<a/2$), $P_{y}$ is odd function about the mirror plane at $x^{\prime}=-a/4$ ($x^{\prime}=a/4$), so that the integral of $\langle P_{y}\rangle$ equates to zero. Thus, the radiation field is zero. These inference were confirmed by numerical calculation of $\langle P_{y}\rangle$, $\langle xP_{y}\rangle$ and $\langle zP_{y}\rangle$ of the leaky resonant mode and BIC. The evanescent field of the BIC in the region with $|z|>h/2$ is approximately equal to $e^{(|z|-h/2)/\delta_{z}}$, with $\delta_{z}$ being the penetration depth into the vacuum. Because the BIC is mainly consisted of superposition of two waveguide modes with propagating constants being $\pm2\pi/a$, $\delta_{z}$ is approximately equal to $\delta_{z}^{BIC}=a/(2\pi\sqrt{1-af_{BIC}/c})$ with $f_{BIC}$ being the eigen frequency of the BIC. By fitting the evanescent field with the numerical result, $\delta_{z}$ of the BICs in Fig. \ref{figure_trans}(d) and (f) are around 220 nm, which is slightly larger than the corresponding theoretical value $\delta_{z}^{BIC}$.

\section{Bistability by tuning incident power}

\begin{figure*}[tbp]
\scalebox{0.63}{\includegraphics{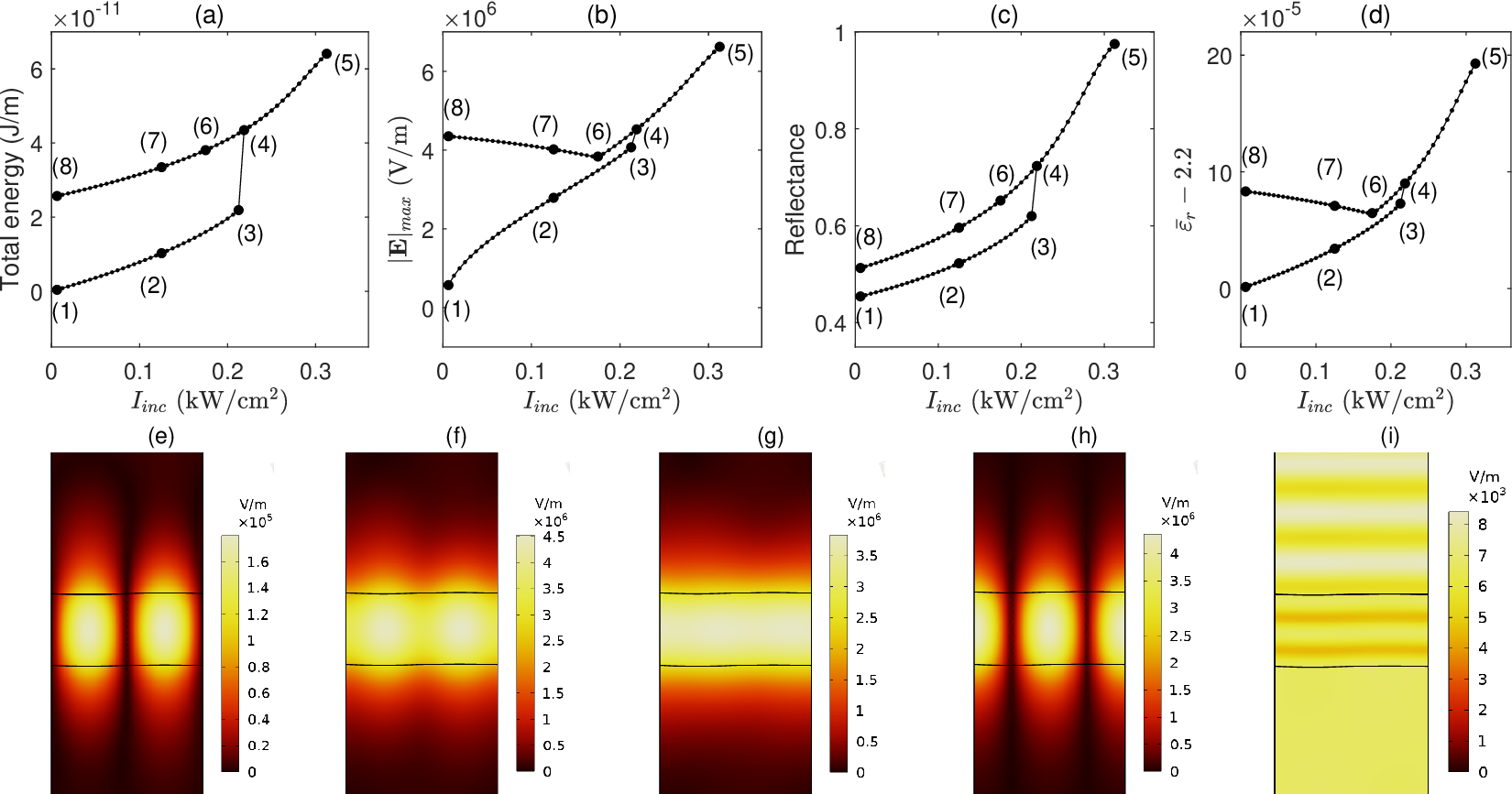}}% Here is how to import EPS art
\caption{ The hysteresis loop of the optical bistable operation of the wavy dielectric grating with $w=3$ nm. The normally incident optical plane wave has out-of-plane polarization, whose frequency is $fa/c=0.774255$. The parameters of the incident power variation are $I_{0}=6.25\times10^{-4}$ kW/cm$^{2}$, $\Delta I=6.25\times10^{-3}$ kW/cm$^{2}$ and $N=50$. The evolution of the (a) total energy in the computational domain, (b) maximum magnitude of the electric field, (c) reflectance, (d) averaged permittivity in the dielectric slab are plotted as small solid dotted lines. Eight selected steps along the hysteresis loop are marked by large solid dots with serial number (1-8) with the same sequence as the direction of the evolution. The field pattern of $|\mathbf{E}|$ at the steps (1), (4), (6), and (8) are plotted in (e), (f), (g), and (h), respectively. The field pattern of $|\mathbf{E}|$ for the erasing step with double frequency and incident power being $I_{0}$ is plotted in (i).   }
\label{figure_loop}
\end{figure*}

In this section, we study the bistability of the optical response, when $I_{inc}$ is adiabatically increased and then decreased. The first step of the simulation starts with a frequency domain calculation with ultra-low $I_{inc}$, designated as $I_{0}$, so that Kerr nonlinearity can be neglected. The subsequent steps of the simulation are the frequency domain calculations with $I_{inc}$ being increased with a small value $\Delta I$ in each step. In each step, the initial value of the electromagnetic field is the convergent solution of the previous step, and the self-consistent solution is obtained through numerical iteration. After $N$ steps, $I_{inc}$ reaches a maximum value $I_{max}=I_{0}+N\Delta I$. For the other $N$ subsequent steps, $I_{inc}$ was decreased by $\Delta I$ in each step. After the simulation of the $2N+1$ steps, $I_{inc}$ returned to the initial value $I_{0}$.

When $I_{max}$ exceeded a certain critical value, the evolution entered a hysteresis loop. An example of $w=3$ nm is shown in Fig. \ref{figure_loop}. The frequency of the incident plane wave is $fa/c=0.774255$, and the corresponding linear reflectance is $0.4542$, which is on the climbing slope of the resonant peak in Fig. \ref{figure_trans}(b). The critical value of $I_{inc}$ was $0.21$ kW/cm$^{2}$. The optical field is characterized by the total energy of the entire computational domain,  maximum value of the amplitude of the electric field (designated as $|\mathbf{E}|_{max}$),  reflectance, and  spatially averaged value of the relative permittivity within the dielectric slab (designated as $\bar{\varepsilon}_{r}$), which are plotted in Fig. \ref{figure_loop}(a), (b), (c), and (d), respectively. Eight typical steps are marked in the evolution curve with the indices in sequence.

In the initial step [step (1)], the incident plane wave excited the leaky resonant mode, as shown in Fig. \ref{figure_loop}(e). Because $I_{inc}$ is equal to $6.25\times10^{-4}$ kW/cm$^{2}$ in the initial step, the amplitude of the electric field of the incident plane wave is $2.17\times10^{3}$ V/m. The incident plane wave excites the leaky resonant mode with a significantly high Q factor, such that the resonant optical field in the wavy dielectric slab is enhanced  $82$ times, as shown in Fig. \ref{figure_loop}(e). The field pattern is a superposition of the incident plane wave and the excited leaky resonant mode with a large amplitude, such that the field pattern is nearly the same as that of the leaky resonant mode in Fig. \ref{figure_trans}(c). In steps (2) and (3), the field pattern is the same as that shown in Fig. \ref{figure_loop}(e), except that the amplitude is larger.

In step (4), $I_{inc}$ exceeds the critical value and the field pattern changes sharply, as shown in Fig. \ref{figure_loop}(f). The nodal lines at $x=0$($x=a/2$) are filled with a sizable optical field, implying that the field pattern is a superposition of the incident plane wave,  leaky resonant mode, and  BIC. Thus,  BIC is excited because of the spatial modification of $\varepsilon_{r}$. The sharp change in the field pattern can be characterized by the sharp change of the total energy, $|\mathbf{E}|_{max}$,  reflectance, and $\bar{\varepsilon}_{r}$. When $I_{inc}$ further increases and reaches step (5), the optical field at $x=0$($x=a/2$) remains nearly the same and that at $x=a/4$($x=3a/4$) increases. Meanwhile, the total energy, $|\mathbf{E}|_{max}$, and $\bar{\varepsilon}_{r}$ were nearly proportional to $I_{inc}$. These phenomena imply that the stored energy in the BIC remains the same, and that the stored energy in the leaky resonant mode is proportional to $I_{inc}$. As $I_{inc}$ decreases to be smaller than the critical value, the field pattern does not sharply change back to that of step (3), but remains  similar to that of step (4) because the stored energy in the BIC locks the spatial pattern of $\varepsilon_{r}$.

As $I_{inc}$ reaches step (6), the field pattern becomes nearly uniform along x-direction, as shown in Fig. \ref{figure_loop}(g), which implies that the amplitudes of the leaky resonant mode and BIC are nearly the same. As $I_{inc}$ further decreases and reaches steps (7) and (8), the stored energy in the BIC is larger than that in the leaky resonant mode, so that the total energy continues to decrease, while $|\mathbf{E}|_{max}$ and $\bar{\varepsilon}_{r}$ hardly change. In step (8), the energy in the leaky resonant mode is much smaller than that in the BIC. Therefore, the field pattern is nearly the same as that of the BIC in Fig. \ref{figure_trans}(d), as shown in Fig. \ref{figure_loop}(h). If $I_{inc}$ is tuned  to  zero, the two-dimensional Helmholtz equation becomes a nonlinear eigenvalue equation. With the spatial distribution of $\mathbf{E}$ and $\varepsilon_{r}$ in step (8) as the initial condition, the self-consistent solution of the nonlinear eigenvalue equation yields a resonant mode, whose field pattern and amplitude are nearly the same as those in Fig. \ref{figure_loop}(h), and Q factor is infinitely large. The eigen-frequency of the resonant mode is slightly different from that of the BIC in Fig. \ref{figure_trans}(c). Thus, the resonant mode is designated as a modified BIC of the wavy dielectric grating with non-uniform $\varepsilon_{r}$. Although $\varepsilon_{r}$ becomes non-uniform, the condition of the symmetry mismatch is preserved; thus, the stability of the BIC is preserved in the modified BIC. At this stage,  optical energy is theoretically stored in the modified BIC without loss. Applying the same incident plane wave by increasing $I_{inc}$ can only drive the system back to step (8-5), but cannot return to the initial state in step (1).

For applications in optical storage devices, in addition to storing optical energy, the ability to release  stored energy is  required. To release the stored energy, the spatial pattern of $\varepsilon_{r}$ has to be changed, which breaks the symmetry mismatch between the modified BIC and the radiative traveling wave. This objective can be achieved through the incidence of plane waves with different frequencies. Using the field pattern in Fig. \ref{figure_loop}(h) and the corresponding $\varepsilon_{r}$ as the initial condition, an iterative solution of the optical field under the incidence of a plane wave with a frequency of $2f$ and $I=I_{0}$ is plotted in Fig. \ref{figure_loop}(i). The amplitude of the field pattern in Fig. \ref{figure_loop}(i) is much smaller than that in Fig. \ref{figure_loop}(h),  implying that the stored energy in the modified BIC is completely released. The field pattern in Fig. \ref{figure_loop}(i) has an interference fringe above the top and bottom surfaces of the wavy dielectric grating, which is due to the interference between the incident  and  reflected fields above each surface. The field pattern below the bottom surface was uniform and mainly consisted of the transmitted plane wave. With the spatial distribution of $\mathbf{E}$ and the corresponding $\varepsilon_{r}$ in Fig. \ref{figure_loop}(i) as the initial condition, the self-consistent solution of the nonlinear eigenvalue equation does not provide any resonant mode with an infinite Q factor near frequency $f$. The numerical results indicate that the incident field with frequency $2f$ destroys the BIC near  frequency $f$. With the field pattern in Fig. \ref{figure_loop}(i) as the initial condition, the iterative solution of the optical field under the incidence of a plane wave with the original frequency $f$ and $I=I_{0}$ is the same as that in step (1). Thus, the system enters another cycle of  bistable operations.

\begin{figure}[tbp]
\scalebox{0.62}{\includegraphics{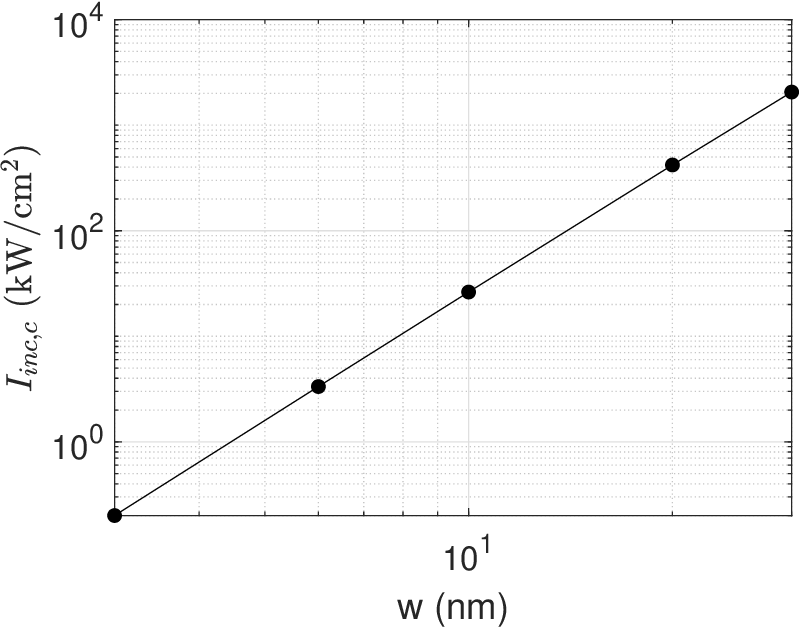}}% Here is how to import EPS art
\caption{ The critical value of $I_{inc}$, where the optical field pattern sharply change, versus $w$ are plotted. For each system with the corresponding $w$, $f$ is chosen from the left part of the resonant peak in Fig. \ref{figure_trans}(b) with the reflectance being 0.46.  }
\label{figure_critical}
\end{figure}

The critical value of $I_{inc}$ between steps (3) and (4), designated as $I_{inc,c}$,  depends on $w$. For a system with a larger $w$, the Q factor of the leaky resonant mode is smaller, and  the enhancement of the optical field within the wavy dielectric grating is weaker. As a result, a larger $I_{inc}$ is required to induce a sufficient modification of $\varepsilon_{r}$. For  systems with varying $w$, $I_{inc,c}$ are plotted in Fig. \ref{figure_critical}. By fitting the data, we found that $I_{inc,c}$ is a quartic polynomial function of $w$ as $I_{inc,c}=w^{4}/403.43$.

\section{Bistability by tuning linearly polarized angle}

\begin{figure}[tbp]
\scalebox{0.58}{\includegraphics{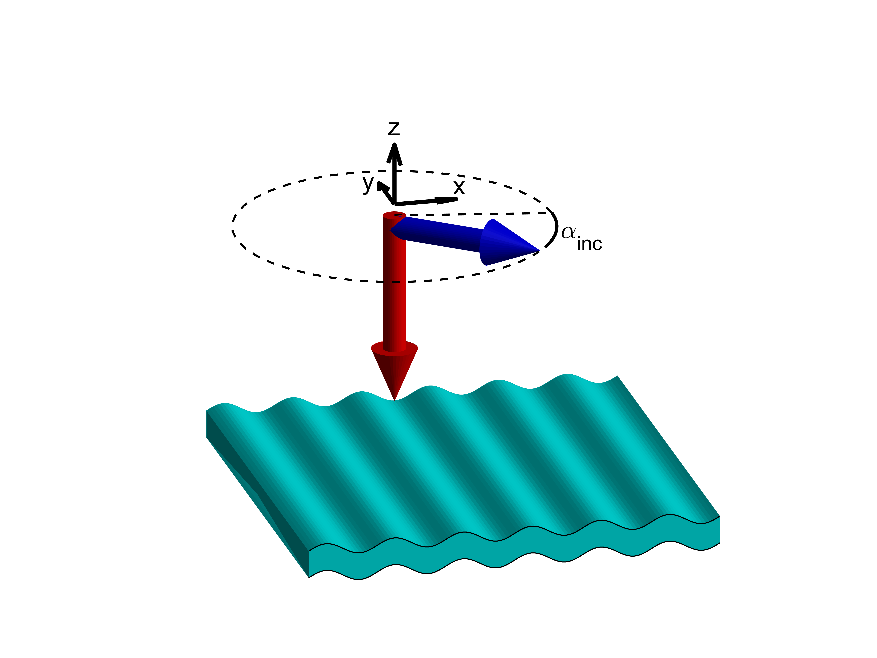}}% Here is how to import EPS art
\caption{ The scheme map of bistability by tuning linearly polarized angle $\alpha_{in}$.  }
\label{figure_PolariScheme}
\end{figure}

\begin{figure}[tbp]
\scalebox{0.64}{\includegraphics{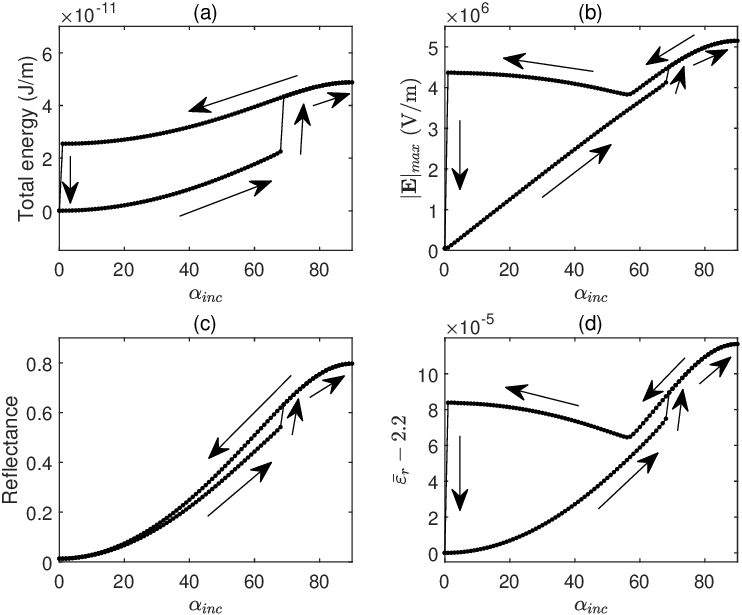}}% Here is how to import EPS art
\caption{ The hysteresis loop of the optical bistable operation of the wavy dielectric grating with $w=3$ nm. The frequency of the normally incident plane optical wave is $fa/c=0.774255$. The incident power is fixed to be $0.3125$ kW/cm$^{2}$. The linear polarization angle firstly increases from 0$^{o}$ to 90$^{o}$ with step size being 1$^{o}$, and then decreases back to 0$^{o}$, with 0$^{o}$ (90$^{o}$) representing in-plane (out-of-plane) linear polarization. Along the hysteresis loop, the (a) the total energy in the computational domain, (b) $|\mathbf{E}|_{max}$, (c) reflectance, (d) $\bar{\varepsilon}_{r}$, versus the linear polarization angle are plotted as small dotted lines. The arrows along the hysteresis loop guide the direction of the bistable operation.  }
\label{figure_loopPolari}
\end{figure}

The hysteresis loop can also be driven by the variation in the linearly polarized angle, which is designated as $\alpha_{inc}$, while keeping the incident power of the normally incident plane wave unchanged. Mechanism of  bistable operation is the same as that in Fig. \ref{figure_loop}, except that the incident power is replaced with the effective incident power. The frequencies of the resonant modes of  TE and TM polarizations are generally different. When the frequency of the incident plane wave is chosen to be at the middle of the climbing slope of the TE-polarized resonant peak in Fig. \ref{figure_trans}(b), the frequency is far  from any TM-polarized resonant peak. Thus, at the same frequency, the TM-polarized incident plane wave with incident power being $I_{inc}=I_{max}$ does not induce a sufficient modification of $\varepsilon_{r}$ for a nonlinear response. The electric field of the incident plane wave can be designated as $\mathbf{E}_{inc}=(\cos\alpha_{inc}\hat{x}+\sin\alpha_{inc}\hat{y})E_{inc}e^{-i2\pi fz/c}$, such that $\alpha_{inc}=0^{o}$ and $90^{0}$  correspond to the TM- and TE-polarized incident plane waves, respectively. For a system with the same parameters as those in Fig. \ref{figure_loop} and $I_{inc}=0.3125$ kW/cm$^{2}$, the evolution of the characteristic parameters versus $\alpha_{inc}$ is plotted in Fig. \ref{figure_loopPolari}. The effective incident power for each $\alpha_{inc}$ can be defined as the incident power of the TE component of the incident plane wave, that is,  $I_{eff}=I_{inc}\sin^{2}\alpha_{inc}$. The critical angle at which the optical field pattern changes sharply is $\alpha_{inc}=68.5^{0}$,  where $I_{eff}=0.2705$ kW/cm$^{2}$. The critical value of $I_{eff}$ is slightly larger than the critical value of $I_{inc}$ in Fig. \ref{figure_loop} because the modification of $\varepsilon_{r}$ by the TM component of the optical field suppresses the excitation of  BIC. After $\alpha_{inc}$ exceeded the critical angle, the evolution of the bistable state was similar to that shown in Fig. \ref{figure_loop}(a-d). As $\alpha_{inc}$ reaches zero, the field pattern sharply changes to be the same as the initial state because the incident TM polarized plane wave releases the stored energy in the modified BIC. Consequently, a complete hysteresis loop is obtained by tuning only $\alpha_{inc}$ without changing the frequency and incident power.

\section{Conclusion}

In conclusion, the bistable processes of the optical field in a wavy dielectric grating consisting of a material with Kerr nonlinearity under normally incident plane waves were studied. For an incident field with TE polarization, by selecting the frequency at the climbing slope of the line shape of the leaky resonant mode and tuning the incident power, the BIC with a frequency near  the leaky resonant mode is excited. After exciting the BIC and tuning off the incident field, optical energy is stored in the system without loss. The incident of a weak optical field with different frequencies or polarizations can release the stored energy and switch the system back to its initial state. For an incident field with a fixed incident power, the system could enter a hysteresis loop by tuning the linear polarization angle. Bistable operation can be applied to the design of optical storage devices and  switches.

\begin{acknowledgments}
This project is supported by the Natural Science Foundation of Guangdong Province of China (Grant No.
2022A1515011578),  the Special Projects in Key Fields of Ordinary Universities in Guangdong Province(New Generation Information Technology, Grant No. 2023ZDZX1007), the Project of Educational Commission of Guangdong Province of China (Grant No. 2021KTSCX064), and the Startup Grant at Guangdong Polytechnic Normal University (Grant No. 2021SDKYA117).
\end{acknowledgments}

\clearpage

\end{document}